\documentclass[fleqn,twoside]{article}
\usepackage{espcrc2,epsfig}

\def\bc{\begin{center}}
\def\ec{\end{center}}
\newcommand{\be}{\begin{equation}}
\newcommand{\ee}{\end{equation}}

\newcommand{\MSbar}{\overline{\mathrm{MS}}}

\newcommand{\nn}{\nonumber}

\newcommand{\gev}{\mathrm{GeV}}
\newcommand{\mev}{\mathrm{MeV}}

\newcommand{\beq}{\begin{equation}}
\newcommand{\eeq}{\end{equation}}
\newcommand{\beqs}{\begin{equation*}}
\newcommand{\eeqs}{\end{equation*}}
\newcommand{\beqn}{\begin{eqnarray}}
\newcommand{\eeqn}{\end{eqnarray}}
\newcommand{\bea}{\begin{eqnarray}}
\newcommand{\eea}{\end{eqnarray}}
\newcommand{\beqns}{\begin{eqnarray*}}
\newcommand{\eeqns}{\end{eqnarray*}}

\def\vdir{v\kern-5.75pt\raise0.15ex\hbox{${\scriptstyle /}$}}
\def\pdir{p\kern-7.8pt\raise0.2ex\hbox{\Big{/}}}
\def\ddir{D\kern-7pt\raise0.2ex\hbox{\big{/}}}
\def\partdir{\partial\kern-7.6pt\raise0.25ex\hbox{/}}
\def\ddirp{D_{\kern-2.75pt\perp}\kern-11pt\raise0.2ex\hbox{\big{/}}\kern+4.5pt}

\newcommand{\AmS}{{\protect\the\textfont2
  A\kern-.1667em\lower.5ex\hbox{M}\kern-.125emS}}

\title{Light Hadron Spectrum, Renormalization Constants and Light Quark Masses with
Two Dynamical Fermions
\thanks{Presented by C.~Tarantino at ``Lattice 2004", Fermilab.}}

\author{$\mathrm{SPQcdR}$ Collaboration\\
D. Be\'cirevi\'c\address[orsay]{Lab. de Physique Th\'eorique (B\^at.210), 
Univ. Paris XI-Sud, Centre d'Orsay, 91405 Orsay-Cedex, France.}, 
P. Boucaud\addressmark[orsay], V. Gim\'enez\address{Dep. de F\'\i sica 
Te\`orica and IFIC, Univ. de Val\`encia, Dr. Moliner 50, E-46100, Burjassot, 
Val\`encia, Spain.}, V. Lubicz\address[diproma3]{Dip. di Fisica, Univ. 
di Roma Tre, Via della Vasca Navale 84, I-00146 Roma, Italy.}\address
[infnroma3]{INFN-Sezione di Roma III, Via della Vasca Navale 84, I-00146 Roma, 
Italy.}, G. Martinelli\address[roma1]{Dip. di Fisica, Univ. di Roma 
``La Sapienza'' and INFN-Sezione di Roma, Piazzale A. Moro 2, I-00185 Roma, 
Italy.}, F. Mescia\addressmark[diproma3]\address[lnf]{INFN, Lab Nazionali 
di Frascati, Via E. Fermi 40, I-00044 Frascati, Italy}\thanks{Partially 
supported by EU IHP-RTN contract HPRN-CT-2002-00311 (EURIDICE)}, S. Simula\addressmark
[infnroma3], and C. Tarantino\addressmark[diproma3]\addressmark[infnroma3]}
       
\begin{document}

\begin{abstract}
The results of a preliminary partially quenched ($N_f=2$) study of the light 
hadron spectrum, renormalization constants and light quark masses are presented.
Numerical simulations are carried out with the LL-SSOR preconditioned Hybrid
Monte Carlo with two degenerate dynamical fermions, using the plaquette gauge
action and the Wilson quark action at $\beta = 5.8$.
Finite volume effects have been investigated employing two lattice volumes:
$16^3 \times 48$ and $24^3 \times 48$.
Configurations have been generated at four values of the sea quark mass
corresponding to $M_{PS}/M_V \simeq 0.6 \div 0.8$.

\vspace{1pc}
\end{abstract}

\maketitle

Lattice QCD calculations of the light hadron spectrum and quark masses have
significantly improved in recent years.
In the quenched approximation, i.e. ignoring quark vacuum polarization effects,
the accuracy has become smaller than $10$\% and the quenching error remains then
the main source of uncertainty.
In the present work we explore sea quark effects in light hadron and quark 
masses, with two dynamical flavors, using the plaquette gauge action and the 
Wilson quark action at $\beta = 5.8$.

Configurations have been generated by using the LL-SSOR preconditioned Hybrid Monte
Carlo~\cite{Duane:de,Gottlieb:mq}.
We have implemented the Leap-Frog integration scheme, with trajectory length 
equal to one and time step $\delta t=4 \cdot 10^{-3}$, and the BiCGStab 
inversion algorithm with iterated residual $r=10^{-15}$.
The acceptance probability is found to be larger than $80$\% and the inversion
rounding error results to be well under control, being of the order of
$10^{-8}$.
We have verified that reversibility is satisfied at the relative level of
$10^{-11}$.

By looking at the autocorrelation times of the plaquette and of the pseudoscalar and
vector two point correlation functions, we made the conservative choice of
selecting configurations separated by steps of $45$ trajectories.

We investigated finite size effects by simulating
two lattice volumes: $16^3 \times 48$ ($L \sim 1.1 fm$) and $24^3 \times 48$ ($L
\sim 1.6 fm$).
At the smaller volume $100$ configurations have been generated at each of the 
four values of the sea hopping parameter ($k_{sea}=0.1535,0.1538,0.1540,0.1541$),
 corresponding to $M_{PS}/M_V \simeq 0.6 \div 0.8$.
At the larger volume, where the simulation is still in progress, we have
generated $50$ configurations at $k_{sea}=0.1535,0.1540$ and $25$ configurations
 at $k_{sea}=0.1538,0.1541$.

\section{Pseudoscalar and vector meson masses}
In order to study the light hadron spectrum and quark masses we have calculated
two points pseudoscalar and vector correlation functions with both degenerate
and non-degenerate valence quarks and with valence quark masses equal or
different to the sea quark mass.

We have tried to implement Jacobi smearing~\cite{Allton:1993wc}, but we 
don't find a significant improvement in the final
determination of hadron masses; the main effect is that the ground state 
can be isolated at smaller times.

\begin{figure}[t]
\begin{center}
\epsfxsize5.0cm\epsffile{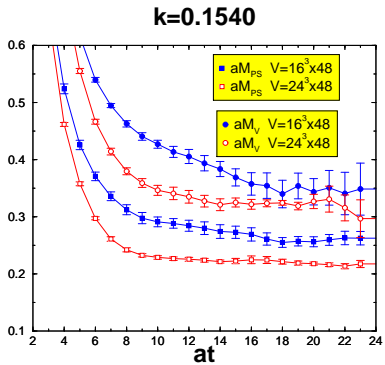}
\end{center}
\vspace*{-1.2cm}
\caption{\label{fig:plateaux}{\sl \small Degenerate
($k_{v1}=k_{v2}=k_{sea}=0.1540$) 
pseudoscalar and vector effective masses, at the
two simulated lattice volumes: $16^3 \times 48$, $24^3 \times 48$.
}}
\vspace*{-0.8cm}
\end{figure}

\begin{figure}[t]
\hspace*{-0.6cm}
\begin{tabular}{cc}
\epsfxsize4.0cm\epsffile{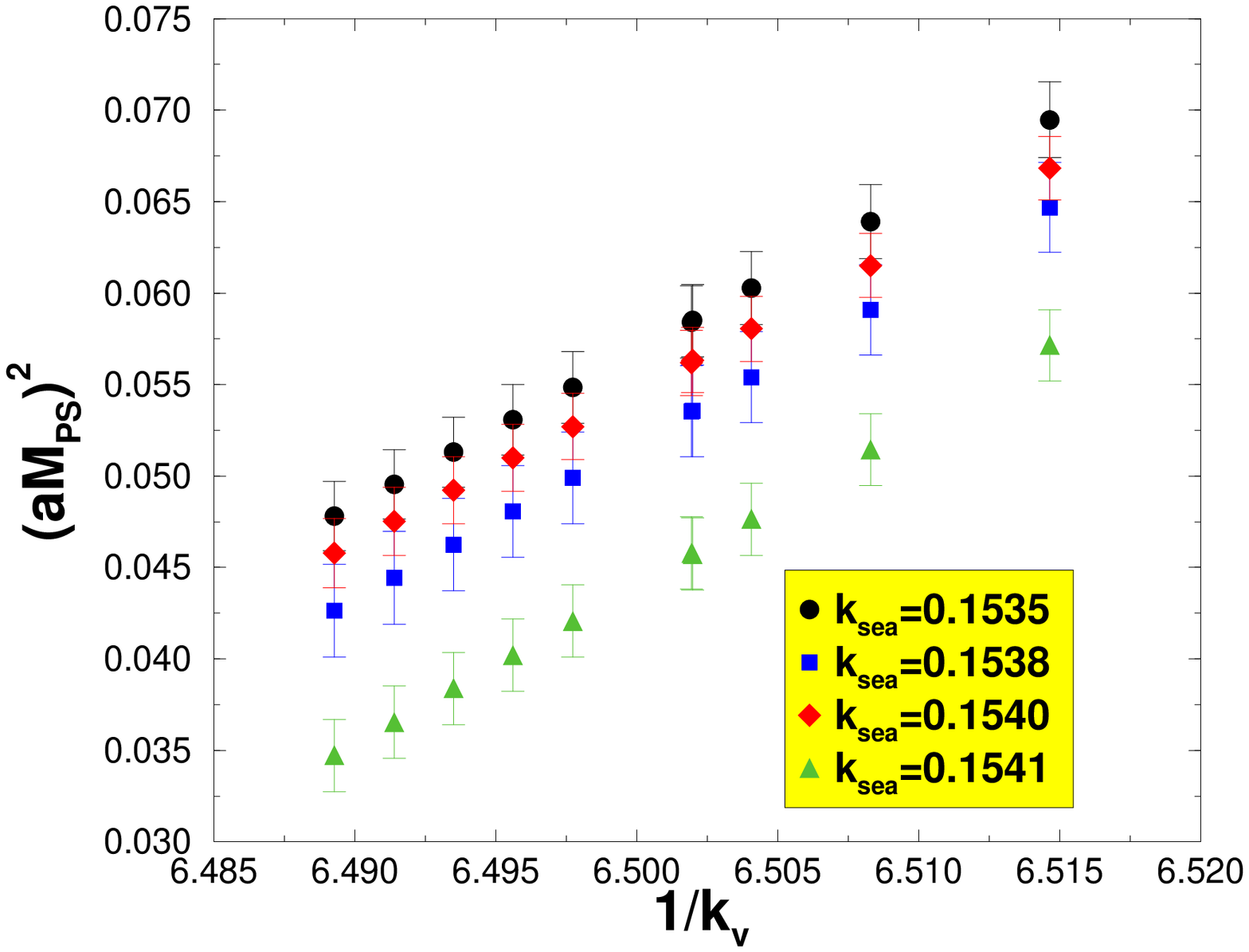} &
\hspace*{-0.5cm}
\epsfxsize4.0cm\epsffile{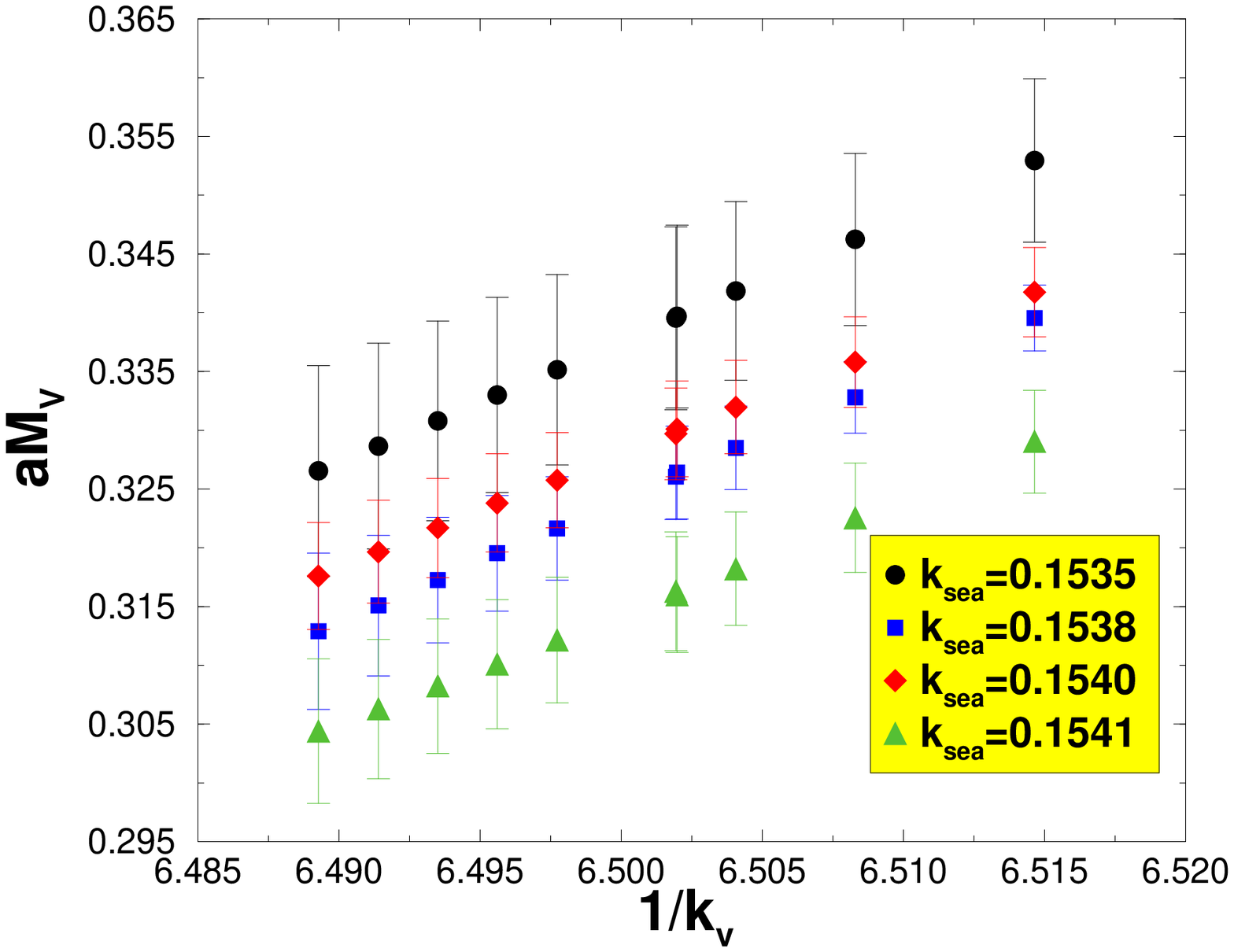} \\
\end{tabular}
\vspace*{-0.9cm}
\caption{\label{fig:diffksea}{\sl \small Pseudoscalar squared(left) and
vector(right) meson masses as a function of the valence quark mass. Different
symbols refer to different values of the sea quark mass.}}
\vspace*{-0.6cm}
\end{figure}

The quality of the plateaux of both pseudoscalar and vector meson effective
masses is shown in Fig.\ref{fig:plateaux}.
One can also see from the plot that finite volume effects are found to be at the
level of $\sim 20$\% on the smaller volume $V=16^3 \times 48$.
Such large effects are perhaps unexpected since they should scale as $\exp{(-M
L)}$, where $M L \simeq 4.4$ for pseudoscalars and $\simeq 5.8$ for vectors in 
our simulation on the small lattice.
We believe that this point requires further investigations.
For the time being, the results we present below follow from the analysis 
performed at the larger volume $V=24^3 \times 48$.
We observe a smooth enhancement of finite volume effects as
the sea quark mass decreases.
In Fig.\ref{fig:diffksea} left(right) we show the pseudoscalar(vector) meson masses as a
function of the valence quark mass ($1/k_v=1/2(1/k_{v1}+1/k_{v2})$).
Different symbols refer to different $k_{sea}$ values.
Within the statistical accuracy, meson masses are found to be linear in the 
valence quark mass, while the dependence on the sea quark mass is less clear.
In particular both pseudoscalar and vector meson masses obtained 
at $k_{sea}=0.1538$ are smaller than those at $k_{sea}=0.1540$, although
compatible within the errors.
As a consequence, the dependence on the sea quark mass is not easily taken into
account in the chiral extrapolations.

In order to determine the lattice spacing and the values of the bare light quark
masses we fit the vector meson masses
linearly in the squared pseudoscalar masses.
Following the method of ``lattice physical planes"~\cite{Allton:1996yv}, we set
the ratio $M_K/M_{K^*}$ equal to its experimental value, thus obtaining for the 
inverse lattice spacing the estimate $a^{-1}=3.0(1)\gev$.
For the pseudoscalar and vector meson masses we find $M_{\pi}=143(3) \mev$, 
$M_{\rho}=805(19) \mev$ (with $M_{\pi}/M_{\rho}$ fixed to its experimental
value) and $M_{\phi}=983(19) \mev$.
 
\section{Renormalization Constants}
The determination of the renormalization constants has been performed
non-perturbatively  by using the RI-MOM method\footnote{For a more detailed explanation see
ref.~\cite{Becirevic:2004ny} and references therein.}.
We have considered the bilinear quark operators $\mathcal{O}_\Gamma = \bar q
\Gamma q$ with $\Gamma = S,P,V,A,T$ standing respectively for
$I,\gamma_5,\gamma_\mu,\gamma_\mu \gamma_5,$ and $\sigma_{\mu \nu}$.
We also present the result for the renormalization constant of the quark field, 
$Z_q$.

Finite volume effects are found to be smaller than $5$\% for all the
renormalization constants and all the values of the sea quark mass.

\begin{figure}[t]
\begin{center}
\epsfxsize5.0cm\epsffile{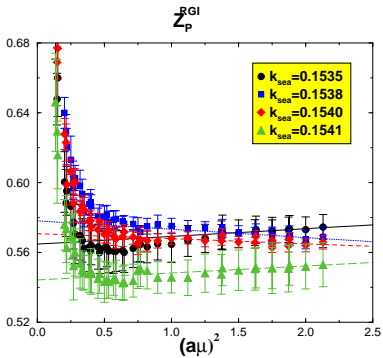} 
\end{center}
\vspace*{-1.0cm}
\caption{\label{fig:zszpksea}{\sl \small $Z_P^{RGI}$ 
at different values of the sea quark mass, as a function of the renormalization
scale $(a \mu)^2$.
At large scales discretization effects introduce a smooth linear dependence on $(a \mu)^2$.}}
\vspace*{-0.6cm}
\end{figure}

The dependence of the renormalization constants on the sea quark mass is weak,
as one can see form Fig.\ref{fig:zszpksea}, where the scale independent
combination $Z_P^{RGI}$~\cite{Becirevic:2004ny} is shown as a function of the
renormalization scale.

The results for the renormalization constants, extrapolated to the chiral
limit, in the RI-MOM scheme at the scale $\mu=a^{-1}$, read
\bea
Z_S=0.72(1)\,, Z_P=0.55(1)\,, Z_V=0.69(2)\,,\nn\\ 
Z_A=0.79(2)\,, Z_T=0.77(5)\,, Z_q=0.79(1)\,.
\eea

\section{Light quark masses}
In order to determine the values of quark masses we used, as 
in ref.~\cite{Becirevic:2002jg}, the standard procedures based
on the vector(V) and axial-vector(A) Ward Identities(WI).

The VWI connects the quark mass renormalization constant to the scalar density
one.
The corresponding mass definition is given by
\be
m_q^{(VWI)}(\mu) = Z_S^{-1}(\mu a)\, \frac{1}{2a} 
\left(\frac{1}{k_q}-\frac{1}{k_{cr}}\right)\,,
\label{eq:mvwi}
\ee
where $k_q$ is the Wilson hopping parameter and $k_{cr}$ is its critical value,
corresponding to $M_{PS}^2(k_v=k_{sea}=k_{cr})=0$.
The AWI, instead, relates the quark mass renormalization constant to
the renormalization constant of the axial and pseudoscalar operators,
leading to the following expression for the renormalized quark mass
\be
m_q^{(AWI)}(\mu) = \frac{Z_A}{Z_P(\mu a)}\,\frac{\langle \sum_{\vec x}
\partial_0 A_0(x) P^{\dagger}(0)\rangle}{2 \langle \sum_{\vec x} P(x) 
P^{\dagger}(0) \rangle}\,.
\label{eq:mawi}
\ee

In order to get the physical values of light quark masses, we study the
dependence of the squared pseudoscalar masses on simulated valence and sea 
quark masses.
The observed behaviour on the sea quark mass doesn't allow us to extract from
the fit the dependence on $(a m_{sea})$, and the dependence on the valence quark
mass is found to be linear.
We perform, therefore, a fit to the form
\be
(a M_{PS})^2 = A + B (a m_{v1}^{(AWI)} + a m_{v2}^{(AWI)})
\label{eq:fitpseudo}
\ee
and a similar expression for the VWI quark mass.
The constant term $A$ in eq.~(\ref{eq:fitpseudo}) is due to 
$\mathcal{O}(a)-$discretization effects and it is only present in the AWI case. 
For the VWI case these effects are automatically included 
in the determination of the critical hopping parameter $k_{cr}$, implying $A=0$.
The physical values of the average up/down ($m_l$) and of the strange
($m_s$) quark masses are then obtained by substituting the experimental pion
and kaon masses on the l.h.s. of eq.~\ref{eq:fitpseudo} and the value of the lattice spacing
($a^{-1}=3.0(1)\gev$).

Quark mass values are converted from RI-MOM at the renormalization scale $\mu =
1/a$ to $\MSbar$ at $\mu=2\gev$ by using RG improved perturbation theory at
$4-$loop accuracy~\cite{Chetyrkin:1999pq}.
The preliminary results read
\bea
m_l^{VWI} = 4.8(5) \mev\,, m_s^{VWI} = 111(6) \mev\,, \nn\\
m_l^{AWI} = 4.5(5) \mev\,, m_s^{AWI} = 103(9) \mev\,.
\label{eq:res}
\eea
Simulations at other values of $\beta$, $k_{sea}$ and larger volumes are required, 
in order to study the continuum limit, the dependence on the sea quark mass and 
finite volume effects.

\vspace*{0.25cm}

It is a pleasure to thank the Lattice-2004 organizers for the pleasant and 
stimulating conference realized at the Fermilab.

\end{document}